\documentclass[9pt]{article}
\usepackage{authblk}
\usepackage{multirow}
\usepackage{booktabs}
\usepackage{upgreek}
\usepackage{graphicx}
\usepackage{dcolumn}
\usepackage{bm}
\usepackage{enumitem}
\usepackage[utf8]{inputenc}
\usepackage[T1]{fontenc}
\usepackage{mathptmx}
\usepackage{etoolbox}
\usepackage[color=lightgray]{todonotes}
\usepackage{siunitx}
\usepackage{amsmath}

\begin{document}

\title{Electron volt energy resolution with ptychography using a broadband continuum spectrum}

\author[a]{Silvia Cipiccia\thanks{corresponding author: s.cipiccia@ucl.ac.uk}}

\author[b]{Wiebe Stolp}
\author[a]{Luca Fardin}
\author[c]{Ralf Ziesche}
\author[c]{Ingo Manke}
\author[b]{Matthieu Boone}
\author[d]{Chris Armstrong}
\author[a]{Alessandro Olivo}
\author[e]{Darren Batey}

\affil[a]{Department of Medical Physics and Biomedical Engineering, University College London, Malet Place Engineering, Gower St, London, WC1E 6BT, United Kingdom}
\affil[b]{UGCT-RP, Department of Physics and Astronomy, Ghent University, Ghent 9000, Belgium}
\affil[d]{Central Laser Facility, Rutherford Appleton Laboratory, Harwell Campus, Didcot OX11 0QX, United Kingdom}
\affil[c]{Helmholtz-Zentrum Berlin für Materialien und Energie Hahn Meitner Platz 1, 14109, Berlin, Germany}
\affil[e]{Diamond Light Source, Harwell Science and InnovationCampus, Fermi
 avenue, Didcot, OX110DE, United Kingdom}

\date{\today}
\maketitle
\begin{abstract}
Ptychography is a scanning coherent diffraction imaging technique successfully applied in the electron, visible and x-ray regimes. One of the distinct features of ptychography with respect to other coherent diffraction techniques is its capability of dealing with partial spatial and temporal coherence via the reconstruction algorithm. 
Here we focus on the temporal and clarify theoretically and with simulations the constraints which affect the energy resolution limits of the ptychographic algorithms. 
Based on this, we design and perform simulations for a broadband ptychography in the hard x-ray regime, which enables an energy resolution down to 1 eV. We benchmark the simulations against experimental ptychographic data from a nickel test sample, by extracting the x-ray absorption near edge spectrum with energy resolution of 5 eV using a continuum spectrum of 20 eV bandwidth. We review the results, discuss the limitations, and provide guidelines for future broadband ptychography experiments, its prospective applications and potential impact on achieving diffraction limited resolutions. 
\end{abstract}

\section*{Introduction}\label{introduction}
Coherent diffraction imaging (CDI) techniques have opened the doors to diffraction limited resolutions by replacing the image forming lens with an algorithm \cite{Gabor_1948, Sayre_1980, Miao_2015,Hoppe_1969} to solve for the phase problem. 
Achieving such high resolution is reliant upon an high quality illumination, namely high coherence. Understanding how to extract the maximum information from the available coherent flux is critical to push the limits of resolution across all imaging regimes. 

The coherence requirements for CDI are explained with clarity in two 2004 papers by Spence \cite{Spence_2004} and van der Veen  \cite{van_der_Veen_2004} in terms of lateral (spatial) and longitudinal (temporal) coherence.
Ptychography \cite{Rodenburg_2007} is a scanning CDI technique, where an object larger than the illumination is scanned across the illumination at overlapping steps while recording diffraction patterns in the far field. By means of scanning, ptychography synthetically extends the lateral coherence length to match the object size. Nellist et al. in the electron regime \cite{Nellist_1995}, and later Thibault and Menzel in the x-ray regime \cite{Thibault_2013}, showed that the additional constraints in the object plane provided by the scanning process, can be used to relax the spatial coherence requirements. If the lateral coherence length is shorter than the illumination width, the partially coherent illumination can be interpreted as a mix of fully coherent states (or modes), incoherent to each other, and each independent mode can be recovered. This approach has enabled ptychography to deal with experimental issues such as large source size or mechanical vibrations in the setup \cite{Li_2016}, making the technique experimentally robust.

Further to the spatial coherence, ptychography has been proven to deal also with partial temporal coherence, due to the presence of multiple energies or broadband radiation \cite{Enders_2014,Yao_2021,Molina_2023,Shearer_2025}, for example by using structured illumination and a multi-kernel iterative deconvolution method \cite { Huixiang_2025}. Dealing with partial temporal coherence was first achieved by Batey et al. in their ptychographic imaging multiplexing (PIM) approach \cite{Batey_2014}, by splitting the broadband illumination into discrete energy channels. Each energy channel is dealt with as an independent measurement and, at the detector plane, all the channels are added together incoherently, that is summing their intensities. Each energy has its own probe (illumination) at the sample, and it is associated to a distinct object. The latter allows to take into account the different response the sample may have to different energies. To give a practical example, let us consider the case where the sample is made out of nickel (Ni) and the illumination contains two discrete energies, one above and one below the Ni k-edge. The object corresponding to the energy channel above the edge will be more absorbing than the object corresponding to the energy channel below the edge. It is worth noting here that the approximation of a single object for multiple energies is possible in the assumption the sample response is constant within the energy bandwidth \cite{Yao_2021}, e.g. far from any absorption edge. 
The PIM approach gives the possibility of relaxing the temporal coherence requirements of standard CDI, as proven in previous works \cite{Yao_2021,PenagosMolina_2023,Schmidt_2022}. However, it has not been investigated yet what determines the ultimate energy separation capability of the algorithm and the implication on the quantitativeness of the technique. For example, is it possible to achieve eV energy separations and enable single acquisition x-ray absorption near edge spectroscopy (XANES) using a broadband spectrum?
In this manuscript we aim to answer these questions theoretically, to confirm with simulations and experiments, and provide guidance for future developments and applications of multi-energy ptychography.  
\section*{Reciprocal and real space constraints for energy resolving ptychographic imaging: theory and simulations}\label{section1}
\begin{figure}
\centering
\includegraphics[scale=0.6]{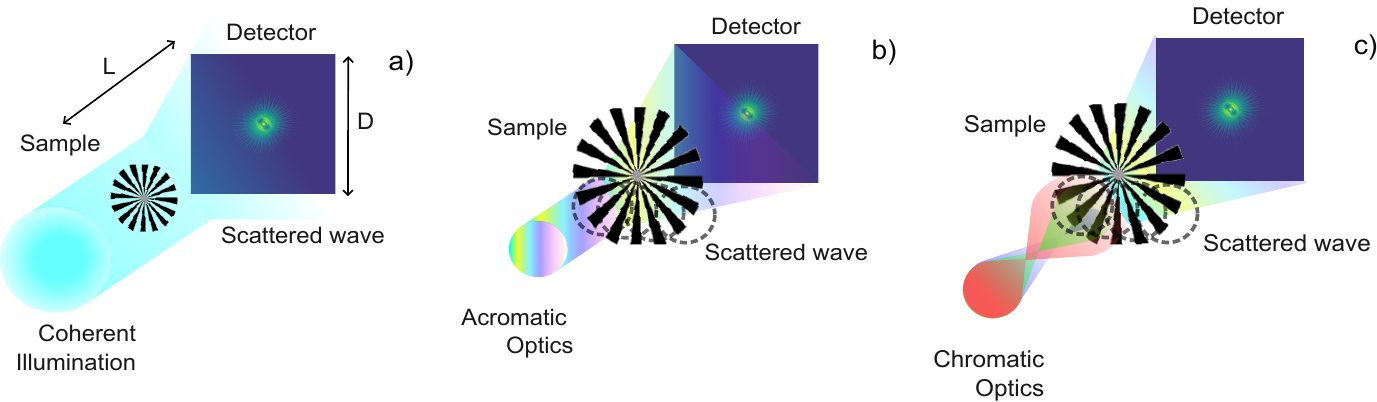}
\caption{a) Standard CDI setup. A coherent illumination is diffracted by the sample. A detector at distance L records the diffraction patterns. b) Achromatic ptychography setup. An achromatic optics forms an energy independent illumination at the sample. c) Chromatic ptychography setup. A chromatic optics forms an energy dependent illumination at the sample. In b) and c)  the sample is scanned through the illumination while a detector records the diffraction patterns.}
\label{fig:CDI_ptycho}
\end{figure}

The work presented in this manuscript is built upon the studies by Spence and van der Veen: in the following section we report the key definitions and formulas without delving into the derivation, which can be found in references \cite{Spence_2004, van_der_Veen_2004}. 
A standard CDI setup is shown in Figure \ref{fig:CDI_ptycho}-a. The lateral (spatial) coherence length $L_c$ at a given distance Z from the source is given by the Van-Cittert-Zernike theorem: $L_c=\lambda Z/S$, where $\lambda$ is the wavelength and S is the source size. For imaging an object of width $W$, the spatial coherence requirement is given by $L_c> 2 W$ \cite{Spence_2004}. The longitudinal (temporal) coherence length $L_t$ depends on the radiation energy E and its bandwidth $\Delta E$: $L_t=\lambda E/\Delta E$. Assuming the spatial coherence requirement is satisfied, the longitudinal coherence limits the best achievable resolution to $d_S$, defined as:
\begin{equation}
d_{S}= \frac{W \Delta E} { E}.
 \label{eqn:d_min_Spence}
\end{equation}

In the conventional CDI sampling condition, the reconstructed pixel size $\Delta x$, depends on the scattering angle subtended by the detector as well as the wavelength of the radiation, and it is given by:
\begin{equation}
\Delta x=\frac{\lambda L}{D}=\frac{\lambda}{\theta} \label{eqn:resolution}
\end{equation}
where L is the sample-detector distance, D the width of the detector active area, and $\theta$ the scattering angle covered by the detector (see Figure \ref{fig:CDI_ptycho}-a). The best resolution achievable in a CDI experiment, $d_{min}$, depends on the maximum angle $\theta_{max}$ at which the diffracted photons are detected; this is affected by multiple factors, e.g. photon statistics, detector noise, scattering properties of the sample, and it is at best $d_{min}=2 \Delta x$, as limited by the experimental geometry.
However,  if the energy bandwidth satisfies the condition:
\begin{equation}
\begin{split}
\frac{\Delta E}{E}> \frac{d_{min} }{W}\label{eqn:ene_con}
\end{split}
\end{equation}
based on the Eq. \ref{eqn:d_min_Spence}, the partial temporal coherence will limit the resolution to $d_{S}>d_{min}$.

In ptychography the multiplexing approach can help to overcome this limit. The energy separation process in PIM is analogous to decoding in cryptography: the energy states are transferred through the data channels (pixels) by encoding through a key, which is the illumination structure, either at the sample or at the detector plane, and the scan position diversity. 

To understand the requirement for PIM to succeed in energy separation, let us start with a ptychography experiment where the illumination contains two energies, $E_1$ and $E_2$, with identical probe size at the sample $W$ (see Figure \ref{fig:CDI_ptycho}-b). Let us assume that $E_1 - E_2 = \Delta E$ satisfies condition (\ref{eqn:ene_con}). 
A feature of size $d_{min}$ diffracts the two energies at positions $z_1$ and $z_2$ respectively in the detector plane, defined by the Bragg law:  

\begin{equation}
z_{1,2}=\frac{\lambda_{1,2}L}{d_{min}}\label{eq:bragg}
\end{equation}
with $\lambda_{1,2}$ the wavelength of $E_{1,2}$.
By rewriting (\ref{eqn:ene_con}) in terms of wavelengths $\lambda_{1,2}$ and using equation (\ref{eq:bragg}), we obtain the following condition:
\begin{equation}
\begin{split}
z_1 - z_2=\frac{(\lambda_1 - \lambda_2)L}{d_{min} }>\frac{L \lambda}{W}=2 p_d \label{eq:spencePtycho}
\end{split}
\end{equation}
where $p_d$ is the detector pixel size in CDI given by:
\begin{equation}
\begin{split}
p_{d}= \Delta \theta_{p} L \\
\Delta \theta_{p}=\frac{\lambda}{2W}\label{eq:detPixel}
\end{split}
\end{equation}
with $\Delta \theta_{p}$ is the angle measured at the sample plane, subtended by a detector pixel.
Equation (\ref{eq:spencePtycho}) tells us that, if condition (\ref{eqn:ene_con}) holds, the photons scattered by a feature of size $d_{min}$ for the two different energies have a separation in the far field larger than two detector pixels: as the information of the two energies is transported in different channels, the energies are expected to be separable by PIM and the resolution limit of $d_S$ to be overcome. Condition (\ref{eq:spencePtycho}) represents the reciprocal space constraint for the energy separation in the PIM algorithm.

The considerations made for the energy separation in the reciprocal space can be extended to the real space: let us rewrite condition (\ref{eqn:ene_con}) in terms of the detector pixel size $p_d$, by using (\ref{eqn:resolution}) and (\ref{eq:detPixel}):
\begin{equation}\label{eq:spencePtychoRS1}
\begin{split}
\frac{\Delta \lambda}{\lambda}=\frac{\Delta E}{E}>2\frac{2p_d}{D}
\end{split}
\end{equation}
After some mathematical substitution (\ref{eq:spencePtychoRS1}) can be rewritten as:
\begin{equation}\label{eq:spencePtychoRS2}
\begin{split}
L\frac{\lambda_1 - \lambda_2}{2 p_d}> \frac{2\lambda L}{D}\\
W_1 - W_2> 2\Delta x\\
\end{split}
\end{equation}
The left side of (\ref{eq:spencePtychoRS2}) can be expressed in terms of  $W_1$ and $W_2$, that is the illumination size at the sample for $E_1$ and $E_2$ respectively. The derived relation tells us that two energies $E_1$ and $E_2$, can be separated by the algorithm, as long as their illumination differs in size at the sample of more than twice the reconstructed pixel size, that is the information of their presence is transported in different channels in the real space. This can be engineered e.g. using chromatic pre-sample optics. Condition (\ref{eq:spencePtychoRS2}) expresses the real space constraint for the energy separation in the PIM algorithm, which is the equivalent to condition (\ref{eq:spencePtycho}) for the reciprocal space.

To demonstrate the  theoretical findings, we performed multi-energies simulations. 
For simplicity, we start the study with a polychromatic illumination made of only two discrete energies and we extend later to multiple energies. To demonstrate the implications of the reciprocal space constraint (condition (\ref{eq:spencePtycho})) we perform a set of simulations with achromatic pre-sample optics, while for the real space constraint (condition (\ref{eq:spencePtychoRS2})) with chromatic pre-focusing optics. To benchmark the theory, we begin with simulations without noise and later we investigate the effect of noise.

The simulations reported in this manuscript are performed using Diffractio \cite{Diffractio_2019}, a python library for diffraction and interference optics. 
To generate the multi-energy datasets, for each energy $E_1$, ... $E_n$ we performed an independent simulation by scanning the object in perfect grid at steps of 1 $\upmu $m, unless otherwise stated. Details regarding the illumination size at the samples are given in the following sections. The diffraction patterns for single energies are combined incoherently at each scanning step, by summing the intensities recorded at the detector plane.
The ptychography reconstructions are performed with an implementation of the ePIE algorithm \cite{Maiden_2009} in PtyREX code \cite{Batey_2014_2}, including the ptychographic multiplexing approach \cite{Batey_2014}. 
To make clear the energy separation, the two sample responses to the different energies are in the first part simulated by generating test targets of different shape (Siemens stars 100\% transmission, 2 rad phase, e.g. see Figure  \ref{fig:pinhole}-A). When investigating the effect of noise on the energy separation and the quantitativeness of the reconstruction, we simulate a Siemens star with same shape and transmission but different phase for different energies. 

The simulated achromatic ptychography setup is shown in Figure \ref{fig:CDI_ptycho}-B). A circular illumination with diameter W= 5$\upmu $m is scanned across a sample. The diffraction pattern is recorded on a detector of width D at a distance L. The illumination at the sample is identical for the two energies $E_1$ and $E_2$ in terms of phase front and modulus, as produced by a pinhole.
$E_1=$ 10 keV is kept constant and $E_2$ is varied between 10.01 keV and 13 keV. The list of simulated energies $E_2$, the corresponding energy difference $\Delta E= E_2 - E_1$, the calculated energy spread limited resolution $d_{S}$ based on equation (\ref{eqn:d_min_Spence}), and the details of the geometry are summarised in Table \ref{table:sim_para}. For the reconstruction of the dual-energy datasets we used as initial guess the probes retrieved from the single energy reconstructions.
The results are shown in Figure \ref{fig:pinhole}. 
 The algorithm separates $E_2= $13 and 11 keV from $E_1=$10  keV (Figure \ref{fig:pinhole} A-B). Some residual mix between the two energies is observed for $E_2=$10.1 keV (Figure \ref{fig:pinhole} C), while the separation fails for $E_2=10.02$ keV (Figure \ref{fig:pinhole} D). Figure \ref{fig:pinhole}-F  present the results in terms of root mean square error (RMSE) between the PIM reconstructions and the single energy reconstructions, by plotting the RMSE as a function of the ratio $d_{S} /\ d_{min}$. This confirms that, the energy separation is successful (low RMSE) when the reciprocal space condition (\ref{eqn:d_min_Spence}) is satisfied, that is $d_{S}/\ d_{min}>1$, and fails otherwise. 
Having simulated an achromatic illumination, there is no difference for $E_1$ and $E_2$ in sample plane (real space constraint, condition (\ref{eq:spencePtychoRS2}) not satisfied). PIM relies only on the modeled difference at the detector plane (reciprocal space constraint, condition (\ref{eq:spencePtycho})).

\begin{table}[t!]
\caption{Simulated $E_2$ energies and corresponding expected resolution limit based on Eq.\ref{eqn:d_min_Spence} and summary of the simulations parameters for the achromatic case.}
\centering
\begin{tabular}{rrr}
$E_2$ [keV] & $\Delta E$ [keV] & $d_{S}$ [nm]\\
\midrule
A) 13.00  & 3 & 1150\\
B) 11.00  & 1 & 455\\
C) 10.10 & 0.1 & 49\\
D) 10.02 & 0.02 & 10\\
E) 10.01 & 0.01 & 5\\
\hline
\multicolumn{2}{l}{Illumination at sample} & 5 $\upmu $m\\
\multicolumn{2}{l}{Scanning step} & 1 $\upmu $m\\
\multicolumn{2}{l}{Sample-detector distance}  & 5 m\\
\multicolumn{2}{l}{Detector pixel size} & 62 $\upmu$m \\
\multicolumn{2}{l}{Detector width} & 256 pixels\\
\multicolumn{2}{l}{$^*$ $d_{min}$} & 76 nm\\
\bottomrule
\end{tabular}\par
\bigskip
$^*$ $d_{min}=2\Delta x$ Calculated for 10 keV, based on Eq.\ref{eqn:resolution}
\label{table:sim_para}
\end{table}

\begin{figure}[t]
\centering
\includegraphics[width=1\linewidth]{Figure2pinholes.pdf}
\caption{A-E) Retrieved phase object for $E_1$ and $E_2$ for different energy separation from 3 keV to 10 eV, as in Table\ref{table:sim_para}. The scale bar in A) corresponds to 1 $\upmu $m and applies all the figures A-E. F) RMSE for the retrieved phase object at $E_1$ between the mixed energies and the monochromatic reconstruction. } 
\label{fig:pinhole}
\end{figure}
To generate diversity in the real space and test the real space constraint (\ref{eq:spencePtychoRS2}), we simulate Fresnel zone plates (FZPs) as pre-sample focusing optics, as their chromatic behaviour is well understood and easily controllable.  The main parameters defining the properties of the optics are the radius $R_{FZP}$, and the outer zone width $\Delta r$, which determine the number of zones $N$ and the focal length $f$ \cite{Attwood_1999},  ( see supplementary material). The chromaticity, that is change in focal length as a function of the energy $\Delta f$, increases linearly with the radius of the lens, hence a larger lens manifests a stronger chromaticity for the same $\Delta r$. 

We performed a set of chromatic simulations for $E_2 =$ 10.01 and 10.001 keV, while keeping $E_1=$ 10 keV, to explore the energy separation not resolved by the achromatic setup. For all the simulations the $E_1$ beam size at the sample is 5 $\upmu $m and the scanning step size is 1 $\upmu $m. The calculated difference in beam size $\Delta W$ for the $E_2$ energies with respect to $E_1$ for four different $R_{FZP}$ are summarised in Table \ref{table:sim_Chorm}. We have varied $d_{min}$ by cropping the detector active area, that is varying the number of pixels used for the reconstruction between $512\times 512$ (full detector) and $256\times 256$ pixels.
\begin{table}[!t]
\caption{Summary of the simulation parameters for the chromatic simulations: difference in probe size at the sample for different FZPs and energies and $d_{min}$.}
\begin{tabular}{c|lc|lc}
$R_{FZP}$ &  \multicolumn{2}{c}{$\Delta E = $10 eV} & \multicolumn{2}{c}{$\Delta E = $1 eV}\\
$[\upmu $m$]$ & $\Delta W$ [nm]& $d_{min}$ [nm] &$\Delta W$ [nm]& $d_{min}$ [nm]\\
\midrule
 $20$ & 1) $45 $& $76$& -.-&  -.- \\
  & 2) $45 $& $38$ &-.-&  -.- \\
\hline
$50$ & 3) $105 $ &  $38$& 5) $10 $ &  $38$\\
\hline
$200 $ & 4) $400$ & $76$& 6) $40$ &  $76$\\
 &   -.-&  -.-& 7) $40 $ & $38$\\
 \hline
$400$ &-.-&  -.- & 8) $ 80 $& $76$ \\
 &-.-&  -.- & 9) $80 $& $38$ \\
\bottomrule
\end{tabular}
\label{table:sim_Chorm}
\end{table}
\begin{figure}[t]
\centering
\includegraphics[scale=0.5]{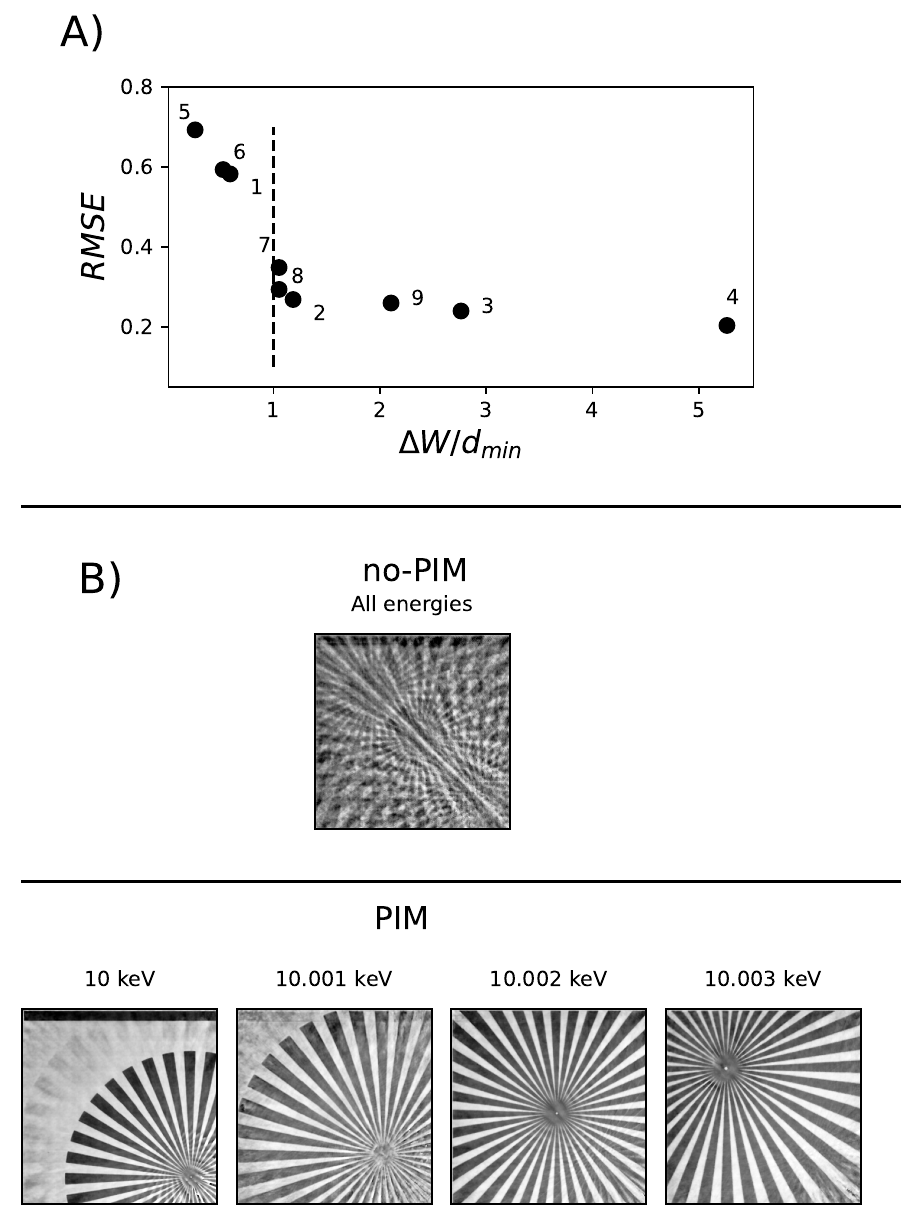}
\caption{A) RMSE for the retrieved object phase at $E_1$ between the single energy and the mixed energies retrieval, as a function of the ratio $\Delta W/d_{min}$. The vertical dotted line defines the limit fo the real space constraint. B) Simulations for four energies. Retrieved phase image using PIM and without PIM. The scale bar corresponds to $1 \mu m$ and applies to all the figures.}
\label{fig:fzp}
\end{figure}
For each entry of Table \ref{table:sim_Chorm} (numbered 1-9) we performed a simulation using the sample-detector distance and detector specifications listed in Table \ref{table:sim_para}. 
The results of the simulations are presented in Figure \ref{fig:fzp}A: the RMSE calculated between the PIM reconstructions and the single energy reconstructions, is plotted as a function of the ratio $\Delta W /\ d_{min}$.
This set of simulations confirms that by using a pre-sample chromatic optics that generate a detectable  difference in probe size at the sample ($> d_{min}$) for different energies the algorithm can separate/decode the energies, even if the reciprocal space constraint (\ref{eq:spencePtycho}) is not satisfied. We have proven so down to 1 eV separation.

Up to here we have simulated a polychromatic beam made of two energies only. The capability of the PIM approach to separate more than two energies has already been proven in previous works \cite{Yao_2021, Batey_2014}. To confirm this is still the case for a very fine energy separation (1eV at 10 keV, hence at $0.01\%$ B.W.), we performed a set of simulations where the diffraction patterns of four different objects corresponding to four energies between 10 and 10.003 keV at steps of 1eV are combined together. The sample is in this case scanned in a $10\times10$ grid at steps of 0.5 $\upmu$m to provide the algorithm enough conditions to solve for all the four complex objects and illuminations. The results are shown in Figure \ref{fig:fzp}-B.

Regarding the effect of noise, in x-ray ptychography photon counting detectors are commonly used to record the diffraction patterns, therefore Poisson noise is generally the main source of noise. To investigate its effect on the energy separation capability of PIM and the quantitativeness of the results, we simulated two energies $E_1 = 8.340$ keV and $ E_2 = 8.350$ keV. For this set of simulations the sample is a Siemens star test pattern with same shape and same transmission at the two energies ($100\%$), but  different phase shift:  $2.0$ and $1.8$ radians respectively at $E_1$ and $E_2$ ($10 \%$ difference). The simulated setup consists of a FZP with 400 $\mu m$ radius and 100 nm outermost width. The sample to detector distance is 5 m, the detector pixel size $74 \mu m$. $256 \times 256$ pixel array, the beam size at the sample is $W_1 = 5.2 \mu m$ and $ W_2 = 4.2 \mu m$. As $\Delta W > 2\Delta x =2 \times 39$ nm, the simulated experimental setup satisfies the real space constraint (\ref{eq:spencePtychoRS2}).

\begin{figure}[t]
\centering
\includegraphics[scale=0.6]{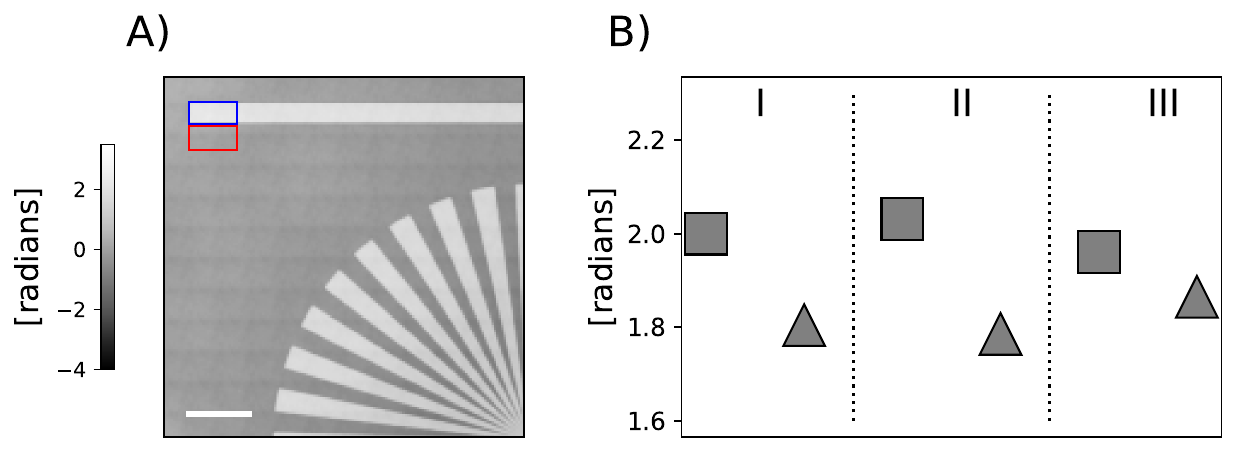}
\caption{A) Retrieved object phase for $E_1 =8.340 keV$ for the monochromatic case. Scale bar, $1\mu m$. B) The retrieve phase is calculated as difference between the average value in blue box and the red box in A), for $E_1$ (square markers) and $E_2$ (triangles) for different noise conditions: I) monochromatic reconstructions with no noise. II) Dual energy reconstruction with Poisson noise for $10^{10} $ photons. III)  Dual energy reconstruction with Poisson noise for $10^8 $ photons.}
\label{fig:noise}
\end{figure}

We performed ptychography retrieval in different noise conditions: I) reference, no noise, separate monochromatic reconstruction; II) PIM with Poisson noise for total flux $10^{10}$ photons; III) PIM with Poisson noise for total flux $10^8 $ photons. The results are summarised in \ref{fig:noise}-B). By increasing the noise level, that is simulating lower statistics, the phase retrieved for the two energies tends to converge to an intermediate value. PIM does not succeed in fully separating the energies which corresponds to a loss of quantitativeness. We also noticed this comes together with the appearance of  pronounced grid artefact in the reconstruction (see Supplementary Figure 2 for the retrieved phase objects of Figure \ref{fig:noise} I-III).

\section*{Experimental results}
\begin{figure}
\centering
\includegraphics[scale=0.6]{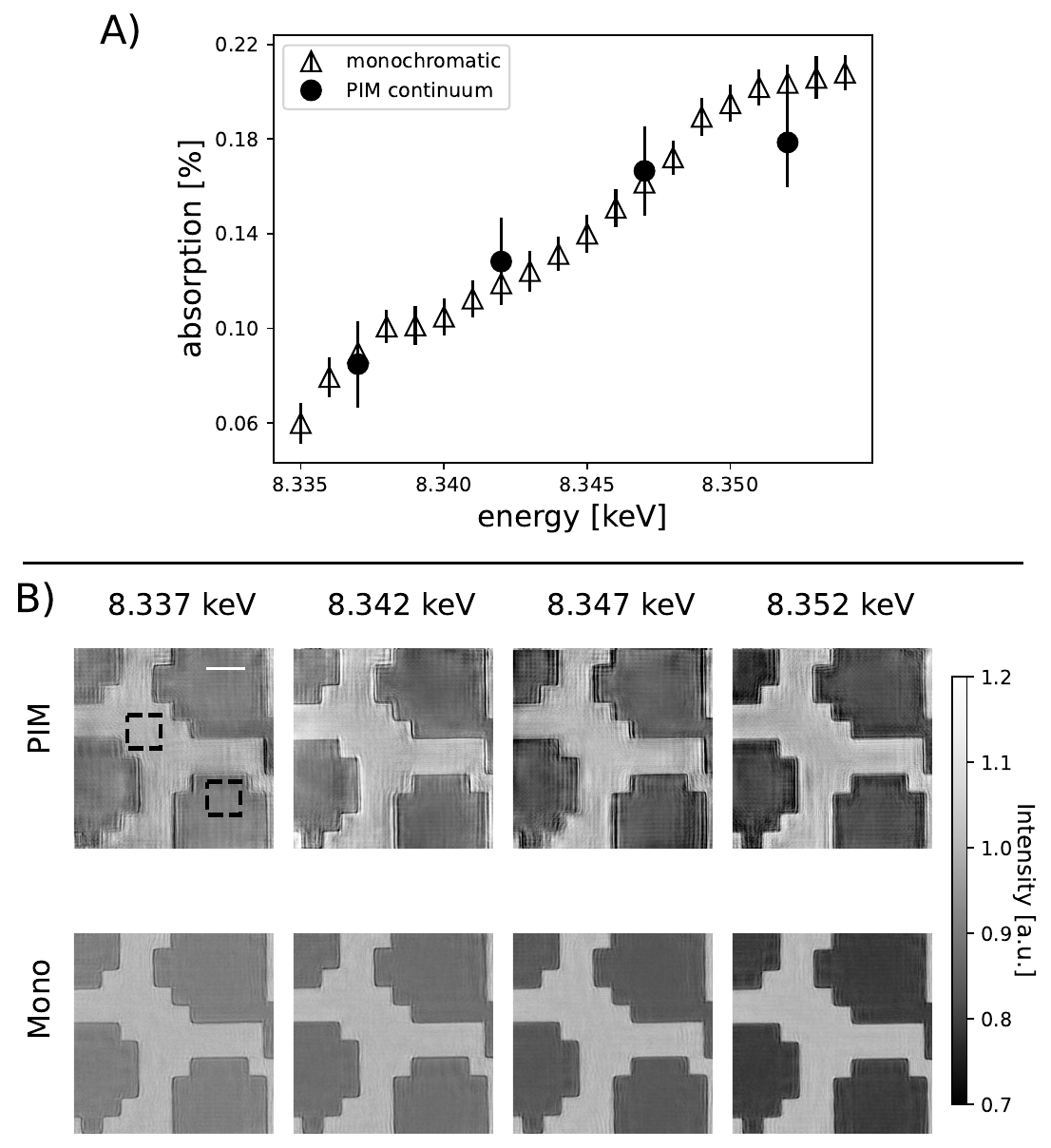}
\caption{A) XANES absorption spectra for the monochromatic and continuum spectrum acquisition. The transmission is calculated as ratio between the average modulus value in the sample and the air within the dashed square shown in B). The error is given by the standard deviation. B) Comparison of the object modulus retrieved for the monochromatic and the continuum spectrum using PIM. White scale bar corresponds to $2\mu m$ and applies to all the images in B).}
\label{fig:experiment}
\end{figure}

To benchmark the theory and simulations we have performed an experiment to image test sample made out or nickel. The experiment was performed at the I13-1 beamline of Diamond Light source\cite{Batey_2018}. The x-ray beam energy was scanned across the Ni k-edge by using a double crystal Si-111 monochromator available at the beamline (bandwidth $1.6\times 10^{-4}$ \cite{Rau2019}). A set of blazed FZP \cite{Gorelick_2011}, 400$\upmu $m radius and 100 nm outer most ring, were used to create a chromatic illumination at the sample. The sample was scanned in a regular grid at steps of 0.25 $\upmu $m, in step-scanning mode \cite{Batey_2022}. The diffraction patterns were recorded 4.8 m downstream the sample with an EIGER 500k photon counting detector (75 $\upmu $m pixel size, $1028 \times 512$ pixels). We acquired twenty separate datasets, corresponding to twenty different energies between 8.335 and 8.354 keV, at step of 1 eV, across the Ni absorption edge at 8.345 keV. In the experimental configuration, the FZP produced a beam size at the sample of approximately 4.9 $\upmu $m at  8.335 keV, corresponding to a  $\Delta W \approx $900 nm every 5 eV. 
We firstly reconstruct the twenty acquisitions separately using 100 iterations of PtyREX code\cite{Batey_2014_2}. 
We estimated the resolution of the image via edge profile (see supplementary material) to be $d_{min}=$ 110 nm, which is smaller than $\Delta W$ and satisfies the real space constraint (\ref{eq:spencePtychoRS2}).
We artificially generated a continuum broadband dataset (see spectrum in Supplementary Figure 1) by summing all the diffraction patterns acquired for the twenty energies at each scanning position. We used the retrieved probes from the single-energy reconstructions as initial guess for the combined energy (PIM) reconstruction of four energies, 5 eV apart from each other. 
The XANES scan obtained with the monochromatic acquisition and that obtained with PIM for the continuum spectrum are compared in Figure \ref{fig:experiment}-A, showing a good agreement.

\section*{Discussion}
We have investigated the limits in energy resolution of the ptychographic imaging multiplexing approach.
We have derived theoretically and benchmarked with simulation the following:
\begin{itemize}
\item[-] for the algorithm to succeed to separate multiple energies, their presence needs to be associated diversity of the illumination, either in the reciprocal space (detector plane, condition (\ref{eq:spencePtycho})) or in the real space (sample plane, condition (\ref{eq:spencePtychoRS2})).
\item[-] The reciprocal and real space constraint (\ref{eq:spencePtycho} and \ref{eq:spencePtychoRS2}) do not need to be satisfied simultaneously to enable the energy separation with PIM. 
\item[-] The real space constraint (\ref{eq:spencePtychoRS2}) tells us that, does not matter how small is the difference between the energies, the algorithm should be able to separate them, as long as a difference in illumination at the sample larger than the resolution of the system, is engineered. 
\end{itemize}

Regarding the chromatic case study, we have used a FZP, which is a common optic used in x-ray ptychography and easy to model within the algorithm. However, the findings can be extended to any other chromatic optics that creates a structured illumination, such in reference\cite{Huixiang_2025}.

In the reported experiment aimed to quantify the senstivity to the energy separation by using a chromatic sample. The broadband was emulated by adding together, at each scanning position, monochromatic diffraction patterns. This process is based on the assumption the scanning positions are identical for the two monochromatic acquisitions. As the acquisitions are sequential the positions are expected to be similar but not identical and we believe this, together with the presence of Poisson noise, prevented achieving a finer energy resolution. Regarding the latter, the total flux on the sample was of the order of $1\times 10^{11}$ photons. The experiment reported here has been performed at a third generation light source: a factor 100 in coherent flux is within reach at fourth generation light sources and this is expect to substantially enhance the energy resolution.  

The study is based on the PIM approach implemented in PtyREX code, yet we do not expect the results to be code dependent and to be extended also to multi-kernel deconvolution approches\cite{Huixiang_2025}. Similarly, we have investigated the broadband ptychography robustness focusing on the x-ray regime, however, we expect the findings to be valid at any  wavelength and radiation. The focus on the x-ray regime is driven by prospective applications for x-ray dichroic and XANES imaging. Ptychography has been proven to be instrumental for magnetic dichroic imaging \cite{Donnelly_2016, Donnelly_2021, Lo_2021} and XANES imaging \cite{Pattammattel_2020, Zhu_2016}. Both techniques require energy scans with steps as small as 1 eV. Based on the findings and the simulations results, 1 eV energy separation is within reach with broadband ptychography, by designing an experimental setup that satisfies the the real space constraint (\ref{eq:spencePtychoRS2}). This would be transformative in terms of reducing the acquisition time and producing inherently co-registered nano scale images across energies. Concerning how to produce a broadband illumination, a multilayer crystal monochromator allows to select bandwidths up to few percent \cite{Boone_2020}. For experimental settings based on undulator sources, the width of the undulator harmonics can be matched to the bandwidth of the multilayer monochromator by tapering the undulator \cite{Mossessian_1995, Batey_2019}.

In laboratory settings \cite{Batey_2021}, because of the lower flux, the role of the monochromator can be replaced with an energy thresholding \cite{Brun_2020} or resolving detector \cite{Batey_2019}. By combining such detectors with an appropriate chromatic optics and the PIM approach, it would be possible to go beyond the energy resolution of the detector itself. 

A final note regarding the potential impact of this study on reaching diffraction limited resolution: recently a 4 nm resolution record has been demonstrated \cite{Aidukas_2024} for hard x-ray ptychography (6.2 keV, 0.02$\%$ B.W.). The diffraction limit is over an another order of magnitude lower. Based on the scaling of flux vs resolution presented by Deng et al. \cite{Deng_2022}, one order of magnitude increase in flux on the sample corresponds to an increase of a factor two in resolution. Therefore, about $10^5$ times more flux than in Ref. \cite{Aidukas_2024} would be required to approach the diffraction limit. Even at fourth generation light sources, this flux enhancement is possibly achievable only by using broader energy bandwidth (few $\%$s), e.g. using a multilayer monochromator. However, the broad bandwidth itself would hinder the resolution. The combination of higher flux, chromatic optics to satisfy the real space constraint (\ref{eq:spencePtychoRS2}), and the PIM reconstruction approach could be the key to unlock diffraction limited x-ray ptychography.

\section*{Acknowledgements}
The Authors thank the Diamond Light Source for the proposal MG30867-1, MG36318-1, MG40558-1, and MG42528-1.We would like to thank XRNanotech for providing the nickel test pattern with very short notice. This work is supported by EPSRC New Investigator Award EP/X020657/1 and Royal Society RGS/R1/231027. W.S and M.B. acknowledge Fonds Wetenschappelijk Onderzoek (FWO-Flanders) for Research Project G030220 and Ghent University Special Research Fund (BOF) for grant BOF/STA/202202/011. A.O. is supported by the Royal Academy of Engineering: CiET1819/2/78.

\bibliographystyle{abbrv}
\bibliography{ArXiv}

\end{document}


\section*{Supplementary Material}

\subsection*{Chromatic simulations }
\begin{figure*}[!ht]
\renewcommand{\figurename}{Supplementary Figure}
\centering
\includegraphics[scale=0.5]{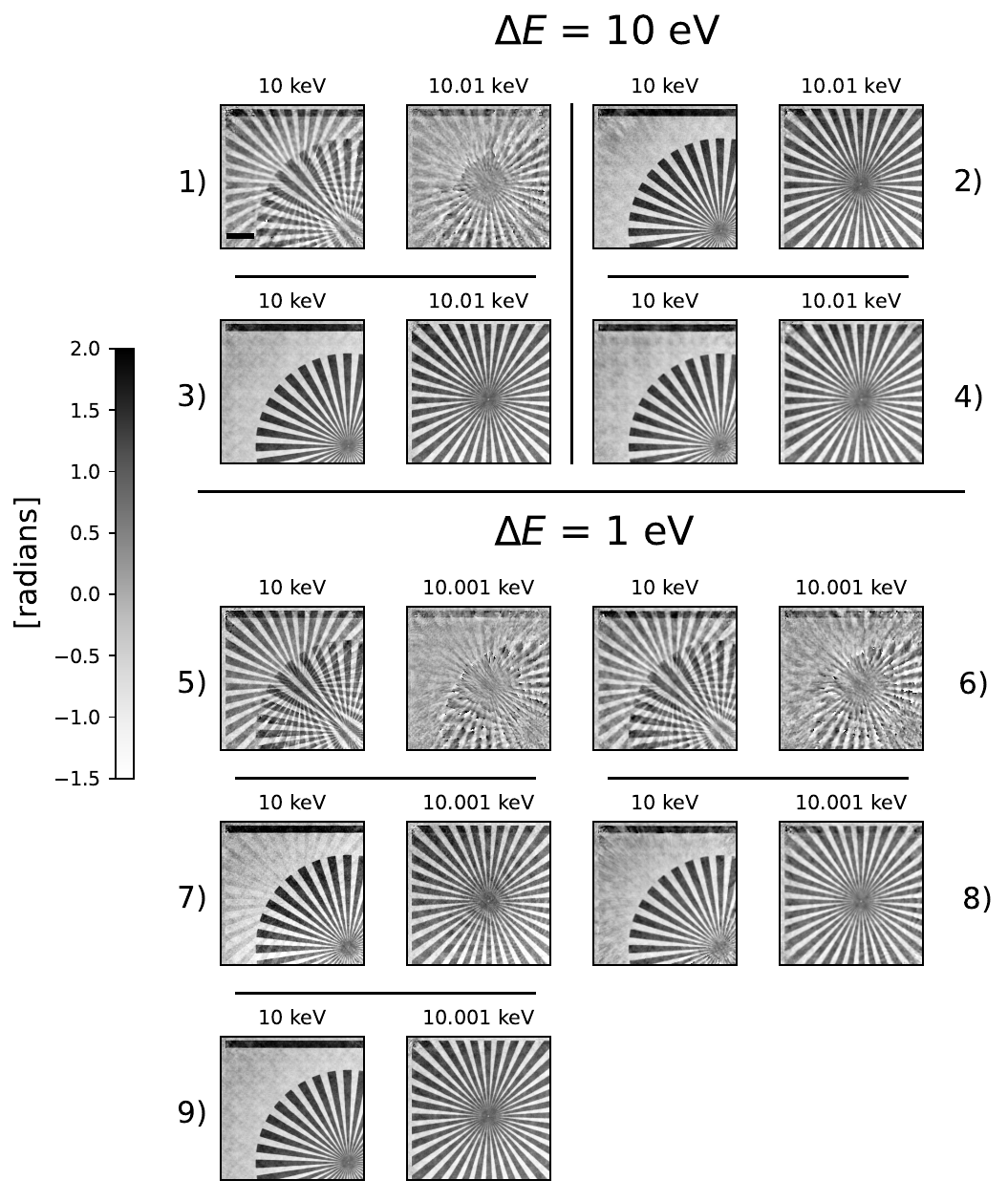}
\caption{ 1-9 PIM Retrieved object phase  for $E_1$ and $E_2$ for different $R_{FZP}$and different $\Delta E$. The scale bar in 1) corresponds to $1 \mu m$ and applies to all the figures.}
\label{fig:fzp}
\end{figure*}

Supplementary Figure \ref{fig:fzp} presents the retrieved phase images for the chromatic pre-focusing optics. The simulations show that, using the 20 $\upmu$m radius FZP, 10 eV separation fails for 256 cropping ($d_{min}=$76 nm), while it is achieved with 512 cropping ($d_{min}=$38 nm). The 10 eV separation is fully achieved with both the 50 and the 200 $\upmu $m radius FZP with $d_{min}=$76 nm and $d_{min}=$38 nm respectively. The 400 $\upmu $m radius FZP successfully separates 1 eV. 

\subsection*{Effect of noise: additional information}
\begin{figure}[h!]
\centering
\includegraphics[scale=0.5]{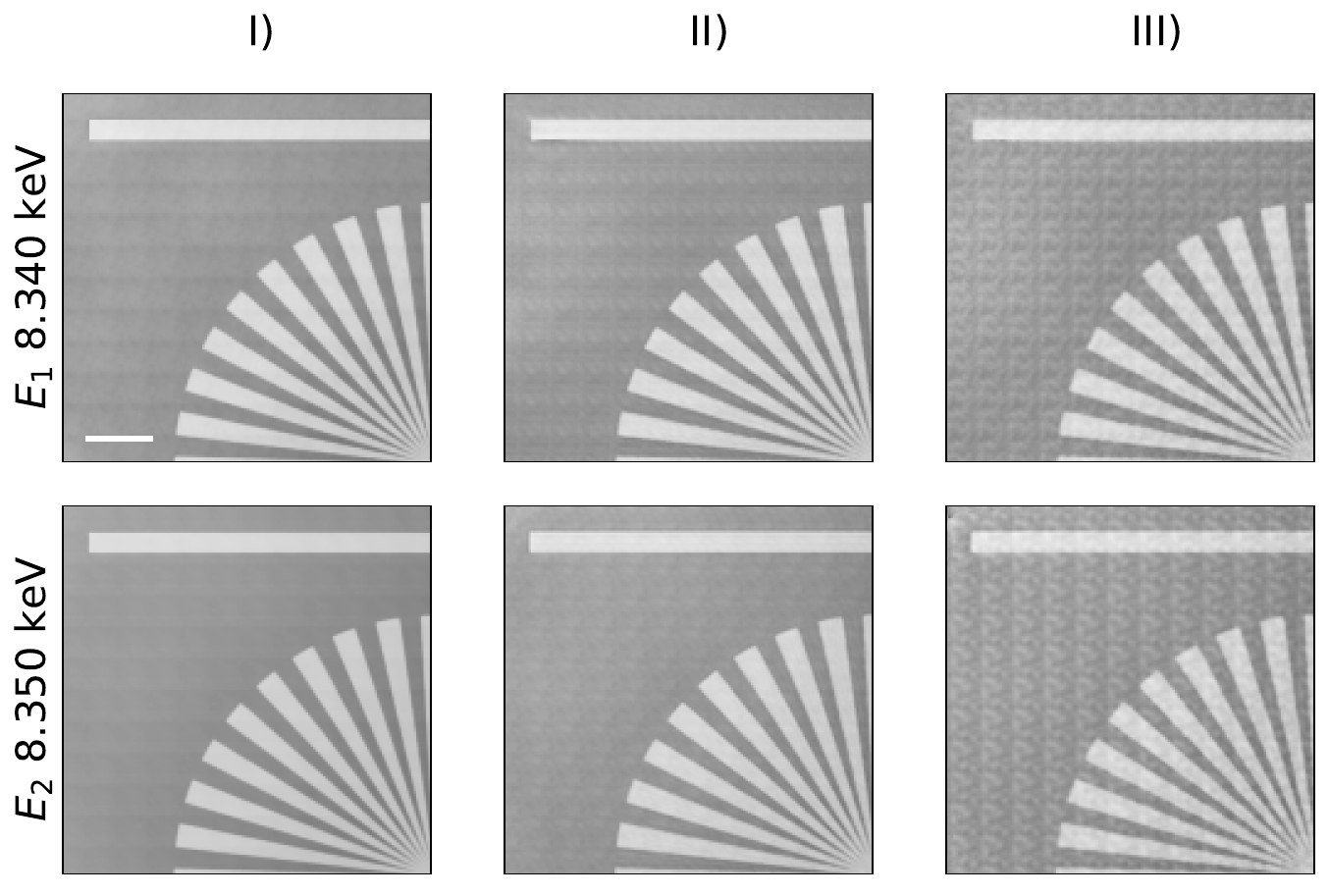}
\caption{Reconstructed phase object for $E_1 =$8.340 keV and $E_2 =$ 8.350 keV for I) monochromatic simulations; II) PIM  with Poisson noise for a total $10^{10}$photons; III) PIM with Poisson noise for a total flux of $10^8$ photons.}
\label{fig:noise_recon}
\end{figure}

The loss in quantitatveness in the energy separation due to an increase of the noise level comes together with more pronounced grid artifact, as shown in \ref{fig:noise_recon}. 

\subsection*{Experimental resolution estimation via edge profile}
\begin{figure}[h!]
\centering
\includegraphics[scale=0.5]{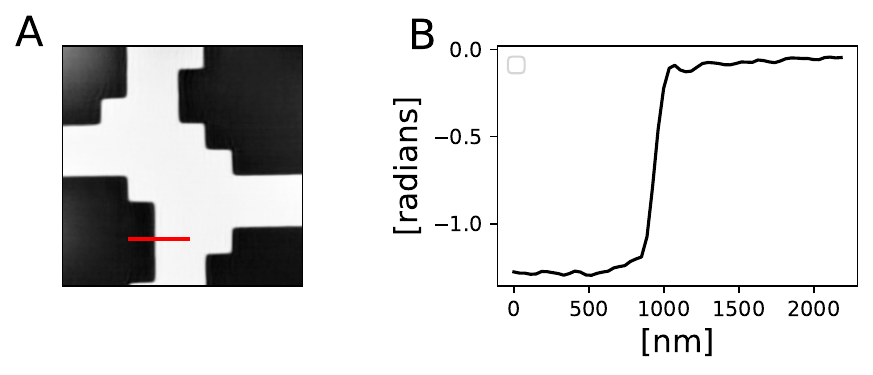}
\caption{A) phase image of the Ni test target from a monochromatic acquisition. B) Edge profile extracted from the red line in A used to estimate the resolution. }
\label{fig:lineprof}
\end{figure}

The phase image from monochromatic acquisition at 8.337 keV was used to estimate the image resolution. From the edge profile extracted from Supplementary Figure \ref{fig:lineprof}-A, a line profile was calculated and the resolution valued as full width half maximum of the line profile, to be 110 nm. see Supplementary Figure \ref{fig:lineprof}.

\subsection*{Synthetic continuum spectrum}
\begin{figure}[h!]
\centering
\includegraphics[scale=0.7]{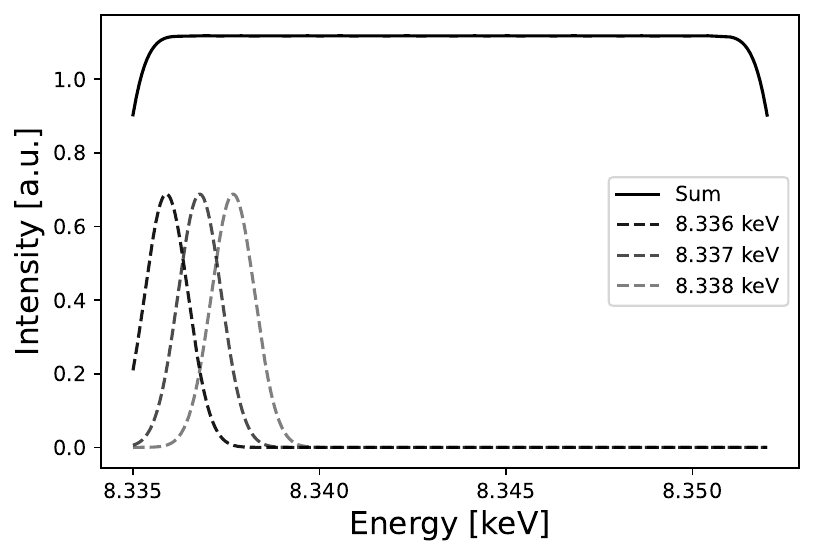}
\caption{ The solid line represent the spectrum obtained by summing the twenty adjacent energy scans. The dashed curves represent three of the twenty gaussian contributing to the final continuum spectrum}
\label{fig:continuumeSpectrum}
\end{figure}

The continuum spectrum was generated by adding up the dataset acquired for the 20 monochromatic acquisition at 1 eV step. The synthetic spectrum produced is shown in \ref{fig:continuumeSpectrum}. This is obtained as sum of twenty gaussian spectra with full width half maximum of $1.6 \times 10^{-4}$\cite{Rau2019}.

\renewcommand{\theequation}{S\arabic{equation}}
\setcounter{equation}{0}
\subsection*{Fresnel zone plate properties}\label{supplementary:FZP}
A Fresnel zone plate is a circular diffraction grating made of concentric alternating opaque and transparent rings, and it is one of the most common chromatic optics used in x-ray ptychography. The main properties of the FZP are summarised by the following equations:
\begin{equation}
N= \frac{R_{FZP}}{ 2\Delta r}  \label{eqn:FZP_Nzones}
\end{equation}
\begin{equation}
f= \frac{4N (\Delta r)^2}{\lambda} \label{eqn:FZP_focus}
\end{equation}
From the above, given a variation in wavelength $\Delta \lambda$ the corresponding change in focal length $\Delta f$ is:
\begin{equation}
\Delta f= - \frac{2R_{FZP} \Delta r}{\lambda^2} \Delta \lambda \label{eqn:FZP_focusVariation}
\end{equation}
where the radius $R_{FZP}$ is the radius, $\Delta r$ the outer zone width, $N$ the number of zones and $f$ the focal length.

\bibliographystyle{abbrv}
\bibliography{ArXiv}